\documentclass[12pt]{article}
\catcode`\@=11
\@addtoreset{equation}{section}

\usepackage{amsbsy,amssymb,latexsym,amsfonts,amsmath,cite}

\allowdisplaybreaks

\global\arraycolsep=2pt
\newcommand\diag{\mathrm{diag}}
\newcommand\CN{\mathcal N}

\newcommand\bref[1]{(\ref{#1})}

\newcommand\CJ{{\mathcal J}}
\newcommand\CI{{\mathcal I}}
\newcommand\CQ{{\mathcal Q}}

\newcommand{\adss}[2]{{AdS$_{#1}\times$S$^{#2}$}}
\newcommand{\ads}[1]{{AdS$_{#1}$}}

\newcommand\BP{\mathbf{P}} 
\newcommand\BJ{\mathbf{J}}

\begin{document}

\begin{flushright}
OIQP-09-07
\end{flushright}

\vspace*{0.5cm}

\begin{center}
{\Large \bf 
Super Galilean conformal algebra
in AdS/CFT
}
\end{center}
\vspace{10mm}

\centerline{\large Makoto Sakaguchi
}

\vspace{8mm}

\begin{center}
{\it Okayama Institute for Quantum Physics \\
1-9-1 Kyoyama, Okayama 700-0015, Japan} \\
{\tt makoto\_sakaguchi\_at\_pref.okayama.jp}
\vspace{5mm}

\end{center}

\vspace{1cm}

\begin{abstract}
Galilean conformal algebra (GCA) is an In\"on\"u-Wigner (IW) contraction of
a conformal algebra,
while
Newton-Hooke string algebra
is an IW contraction of an AdS algebra which is the isometry of an AdS space.
It is shown that the GCA is a boundary realization of the Newton-Hooke string algebra
 in the bulk AdS.
The string lies along the direction transverse to the boundary,
and the worldsheet is \ads{2}.
The one-dimensional conformal symmetry so(2,1)
and rotational symmetry so($d$)
contained in the GCA are realized as
the symmetry on the \ads{2} string worldsheet
and rotational symmetry in the space transverse to the \ads{2} in \ads{d+2},
respectively.
It follows from this correspondence that
32 supersymmetric GCAs
can be derived as IW contractions of
superconformal algebras, psu($2,2|4$),
osp($8|4$)  and osp($8^*|4$).
We also derive 
less supersymmetric GCAs from su($2,2|2$),
 osp($4|4$), osp($2|4$) and osp($8^*|2$).

\end{abstract}

\newpage
\tableofcontents

\section{Introduction}

The AdS/CFT conjecture \cite{AdS/CFT,GKPW}
has attracted much interest
in studies of fundamental aspects of strings
and in
its applications to actual experiments.

One of the main difficulties
to rigorously  prove this conjecture 
is to explicitly quantize the type IIB superstring on \adss{5}{5}
\cite{MT}
(AdS superstring).
So it is still important to look for a limit which extracts
a solvable subsector of the full AdS superstring. 
One such limit is the Penrose limit \cite{Penrose},
which corresponds to taking a close-up of a null geodesic.
Under the limit \cite{Penrose_limit},
the \adss{5}{5} background 
reduces to the pp-wave background \cite{IIBpp}.
The superstring on this background
was exactly solved in the light-cone gauge fixing \cite{MT:ppwave},
and the BMN operator correspondence was found \cite{BMN}.
The symmetry of the latter theory,
the super pp-wave algebra, is an In\"on\"u-Wigner (IW) contraction \cite{IW} of 
the super-isometry of the \adss{5}{5} background,
the super-\adss{5}{5} algebra
\cite{HKS:IW}.

Another limit is the non-relativistic (NR) limit\cite{GGK,BGK}
\footnote{
The NR string theory in flat space was studied in \cite{NR:flat1}\cite{NR:flat2}.},
which corresponds to taking a close-up of an \ads{2} string worldsheet.
The string theory reduces to a free theory of three massive scalars,
five massless scalars and eight massive fermions
propagating on \ads{2} in the static gauge \cite{GGK}
\footnote{
The physical spectrum was obtained in \cite{NRstring:spectrum},
and the semiclassical partition function was evaluated in \cite{DGT}.
}.
These modes correspond to fluctuations of the string worldsheet
\cite{SY:NR}.
The non-normalizable modes
which correspond to fluctuations reaching the boundary
 deform the Wilson loop on the boundary,
and result in operator insertions in the Wilson loop.
The symmetry of the NR string theory
is the super Newton-Hooke string algebra
which is an IW contraction of the super \adss{5}{5} algebra
 \cite{GGK}.
The analysis was extended to supersymmetric branes in \adss{5}{5},
\adss{4}{7} and \adss{7}{4} in \cite{SY:NH}.

\medskip

More recently,
there have been considerable activities
in studies of strongly coupled systems.
The NR conformal field theory (CFT)
which has Schr\"odinger symmetry \cite{Sch} is discussed in \cite{Henkel-MSW,NS},
and is relevant to studies of ultracold atoms at unitarity.
The holographic dual description was proposed in 
\cite{Son,Balasubramanian:2008dm}\cite{Goldberger:2008vg,Barbon:2008bg}
\footnote{
Further studies along this line includes
\cite{SY:Sch}--\cite{Barbon:2009az}.
See \cite{Hartnoll:2009sz} for a review.
}.
The Schr\"odinger symmetry is
a non-relativistic analog of the conformal symmetry
with dynamical exponent $z=2$.
The Schr\"odinger symmetry in $d$ spacial dimensions,
sch($d$), 
is the extension of the Galilean symmetry
generated by time translation $H$, space translation $P_i$,
space rotation $J_{ij}$ and Galilean boost $G_i$,
to include dilatation $D$, Galilean special conformal $K$,
and a center $M$,
where $i=1,\cdots,d$.
The sch($d$) is a subalgebra of the conformal algebra in $(d+2)$-dimensions,
so($d+2,2$),
and thus of the \ads{} algebra in $(d+3)$-dimensions.
From this, super Schr\"odinger algebras were derived
as subalgebras of
superconformal algebras, psu($2,2|4$) and osp($8^{(*)}|4$)
in \cite{SY:Sch,SY:Sch_less}.
Super Schr\"odinger-invariant field theories are obtained as
non-relativistic limits
of relativistic field theories in
\cite{LLM,
NRSY,
Lee:2009mm}.

Galilean conformal algebra (GCA) \cite{Lukierski:2005xy},
which is the main subject in the present paper,
is another non-relativistic analog of the conformal symmetry
with dynamical exponent $z=1$. 
Recent studies on this issue includes
\cite{Bagchi:2009my}--\cite{Bagchi:2009ca}
\cite{Taylor:2008tg}--\cite{Barbon:2009az}.
It is the extension of the Galilean algebra
with dilatation $D$, Galilean special conformal $K$
and acceleration $K_i$. 
It is known that the GCA is an IW contraction of the conformal algebra.
The conformal algebra in $(d+1)$-dimensions is the \ads{d+2} algebra.
So the GCA on the boundary should be related to a certain IW contraction
of the \ads{d+2} algebra.
In fact, 
it is shown that the Newton-Hooke string algebra
\cite{GGK,BGK} is the bulk realization of the GCA
on the boundary.
The string lies along the direction transverse to the boundary.
The GCA has  so($1,2) \times$\,so($d$) as a subalgebra.
The one-dimensional conformal algebra
so($1,2$)\,$\cong$\,sl(2) is 
the symmetry of the \ads{2} worldsheet.
The so($d$) is rotational symmetry
in the space transverse to the \ads{2} in \ads{d+2}.
Taking a close-up of \ads{2} in \ads{d+2}
corresponds the IW contraction of the conformal algebra to
the GCA.

From this observation, we  derive super GCAs from
superconformal algebras in this paper.
Firstly, we derive super GCAs in ($1+3$)-dimensions
as IW contractions
of the four-dimensional superconformal algebras, psu(2,2$|$4)
and su($2,2|2$).
This corresponds to taking a close-up of an \ads{2} string in \adss{5}{5}.
Secondly, super GCAs in $(1+2)$-dimensions
are derived as IW contractions of the three-dimensional superconformal algebras,
osp($8|4$), osp($4|4$) and osp($2|4$).
This corresponds to taking a close-up of an \adss{2}{1} M2-brane in \adss{4}{7}.
Thirdly, 
we derive  super semi-GCAs in $(1+5)$-dimensions
as IW contractions of six-dimensional superconformal algebras,
 osp($8^*|4$) and osp($8^*|2$).
 This corresponds to taking a close-up of an \ads{3} M2-brane in \adss{7}{4}.
The resulting 32 supersymmetric GCAs
are boundary realizations of the super Newton-Hooke algebras
of a brane
derived in \cite{GGK}\cite{SY:NH}.
Less supersymmetric (semi-)GCAs
obtained in this paper give new 
super Newton-Hooke algebras
of a brane. 
Various supersymmetric (semi-)GCAs
can be derived by considering
1/2 BPS brane configurations in \cite{SY:NH},
but we will not complete them here.

\medskip

This paper is organized as follows.
First, we show that 
GCA is the boundary realization of the Newton-Hooke string algebra,
and that
the Newton-Hooke brane algebra is realized as the semi-GCA on the boundary.
In section 3,
super GCAs are derived from the four-dimensional $\CN=4$ and 
$2$ superconformal algebras 
psu($2,2|4$) and
su($2,2|2$), respectively.
In section 4, we derive super GCAs from the three-dimensional $\CN=8$,
$4$ and $2$
superconformal algebras osp($8|4$), osp($4|4$) and osp($2|4$),
respectively.
We also derive super GCAs from the six-dimensional $\CN=4$ and $2$ 
superconformal algebras osp($8^*|4$) and osp($8^*|2$),
in section 5.
The last section is devoted to a summary and discussions. 

\section{GCA from Conformal algebra}

Conformal algebra in $(d+1)$-dimensions is generated by
translation $\tilde P_\mu$,
special conformal transformation $\tilde K_\mu$,
dilatation $\tilde D$
and
Lorentz rotation $\tilde J_{\mu\nu}$,
where $\mu=0,1,\cdots,d$.
The commutation relation is given by
\footnote{We suppress trivial commutators in this paper.}
\begin{eqnarray}
&&
[\tilde J_{\mu\nu},\tilde J_{\rho\sigma}]=\eta_{\nu\rho}\tilde J_{\mu\sigma} +\mbox{3-terms}~,
\nonumber\\&&
[\tilde J_{\mu\nu},\tilde P_\rho]=\eta_{\nu\rho}\tilde P_{\mu}-\eta_{\mu\rho}\tilde P_{\nu}~,~~~
[\tilde J_{\mu\nu},\tilde K_\rho]=\eta_{\nu\rho}\tilde K_{\mu}-\eta_{\mu\rho}\tilde K_{\nu}~,
\nonumber\\&&
[\tilde D,\tilde P_\mu]=\tilde P_\mu~,~~~
[\tilde D,\tilde K_\mu]=-\tilde K_\mu~,
~~~
[\tilde P_\mu,\tilde K_\nu]=\frac{1}{2}\tilde J_{\mu\nu}+\frac{1}{2}\eta_{\mu\nu}\tilde D~,
\label{conformal}
\end{eqnarray}
where $\eta_{\mu\nu}=\diag (-1,+1.\cdots,+1)$.

Let us rename and scale generators as
\footnote{The contraction used in \cite{Lukierski:2005xy}
\begin{eqnarray*}
\tilde P_0=\frac{1}{c}H~,~~
\tilde P_i=P_i~,~~
\tilde K_0=cK~,~~
\tilde K_{i}=c^2 F_i~,~~
\tilde D=D~,~~
\tilde J_{i0}=cK_i~,~~
\tilde J_{ij}=J_{ij}~,~~~
\mbox{and}~~c\to\infty
\end{eqnarray*}
leads to the same result.
}
\begin{eqnarray}
\tilde P_0=H~,~~
\tilde K_0=K~,~~
\tilde D=D~,~~
\tilde J_{ij}=J_{ij}~,~~
\tilde P_i = \omega P_i~,~~
\tilde K_i = \omega  K_i~,~~
\tilde J_{i0} = \omega  G_i~
\label{contraction}
\end{eqnarray}
where $\mu=(0,i)$ and $i=1,\cdots,d$.
Substituting these into \bref{conformal}, 
one sees that the limit $\omega \to\infty$ is
non-singular.
This implies that, under a contraction $\omega \to\infty$, 
the conformal algebra \bref{conformal} reduces to
\begin{eqnarray}
&&
[D,H]=H~,~~
[D,K]=-K~,~~
[H,K]=-\frac{1}{2}D~,~~
\nonumber\\&&
[D,P_i]=P_i~,~~
[D,K_i]=-K_i~,~~
[H,K_i]=-\frac{1}{2}G_i~,~~
\nonumber\\&&
[H,G_i]=P_i~,~~
[K,G_i]=K_i~,~~
[K,P_i]=-\frac{1}{2}G_i~,~~
\nonumber\\
&&
[J_{ij},J_{kl}]=\delta_{jk}J_{il}+\mbox{3-terms}~,~~
[J_{ij},F_k]=\delta_{jk}F_i-\delta_{ik}F_j~,~~
F_i=\{P_i, K_i,G_i\}~.
\label{GCA}
\end{eqnarray}
The resulting algebra
\begin{eqnarray}
\{H,~K,~D,~
J_{ij},~
P_i,~K_i,~G_i
\}
\end{eqnarray}
is the GCA.
The set of generators $\{H,K,D\}$
generates sl($2,R) \cong$ so$(2,1)$
while $\{J_{ij}\}$ generates so($d$).

\subsection{GCA and Newton-Hooke string algebra}

On the other hand, the conformal algebra in $(d+1)$-dimensions
is equivalent to
 the AdS algebra in $(d+2)$-dimensions
\begin{eqnarray}
[\BP_a,\BP_b]=\BJ_{ab}~,~~
[\BJ_{ab},\BP_c]=\eta_{bc}\BP_a-\eta_{ac}\BP_b~,~~
[\BJ_{ab},\BJ_{cd}]=\eta_{bc}\BJ_{ad}+\mbox{3-terms}~,
\end{eqnarray}
where $a=0,1,\cdots,d+1$.
It reduces to the conformal algebra \bref{conformal}
by 
\begin{eqnarray}
\BP_{\mu}=\tilde P_\mu+\tilde K_\mu~,~~
\BP_{d+1}=\tilde D~,~~
\BJ_{\mu\nu}=\tilde J_{\mu\nu}~,~~
\BJ_{\mu d+1}=\tilde K_\mu-\tilde P_\mu~.
\end{eqnarray}
It is easy to see that the scaling \bref{contraction}
corresponds to
\begin{eqnarray}
\BP_{\bar a}\to \BP_{\bar a}~,~~
\BJ_{ij}\to \BJ_{ij}~,~~
\BJ_{\bar a\bar b}\to \BJ_{\bar a\bar b}~,~~
\BP_i\to \omega \BP_i~,~~
\BJ_{i\bar a}\to \omega \BJ_{i\bar a}~,
\label{NH:scaling}
\end{eqnarray}
where $\bar a=0,d+1$.
This is nothing but the contraction which leads to
the Newton-Hooke string algebra
\cite{BGK}\cite{GGK}(see also \cite{SY:NH}).
The string lies along $(d+1)$-th direction.
By the contraction, the AdS algebra reduces to
the Newton-Hooke string algebra
\begin{eqnarray}
&&
[\BP_{\bar a},\BP_{\bar b}]
=\BJ_{\bar a\bar b}~,~~
[\BJ_{\bar a\bar b},\BP_{\bar c}]
=\eta_{\bar b\bar c}\BP_{\bar a}-\eta_{\bar a\bar c}\BP_{\bar b}~,~~
[\BJ_{\bar a\bar b},\BJ_{\bar c\bar d}]
=\eta_{\bar b\bar c}\BJ_{\bar a\bar d} +\mbox{3-tems}~,~~
\nonumber\\&&
[\BP_{\bar a},\BP_{i}]=\BJ_{\bar a i}~,~~
[\BJ_{i\bar a},\BP_{\bar b}]
=\eta_{\bar a\bar b}\BP_{i}~,~~
[\BJ_{\bar a\bar b},\BJ_{i\bar c}]
=\eta_{\bar a\bar c}\BJ_{\bar b i}
-\eta_{\bar b\bar c}\BJ_{\bar a i}~,~~
[\BJ_{ij},\BJ_{k\bar c}]
=\eta_{jk}\BJ_{i\bar c}
-\eta_{ik}\BJ_{j\bar c}~,~~
\nonumber\\&&
[\BJ_{ij},\BP_{k}]=\eta_{jk}\BP_{i}-\eta_{ik}\BP_{j}~,~~
[\BJ_{ij},\BJ_{kl}]
=\eta_{jk}\BJ_{il} +\mbox{3-terms}~.
\label{NH}
\end{eqnarray}
The first three commutation relations
represent \ads{2} symmetry sl($2,R$).
This is the symmetry on the string worldsheet extending along $(0,d+1)$-th directions.
The last commutation relation is so($d$)
which is the rotation in the space transverse 
to the \ads{2} in \ads{d+2}.
The Newton-Hooke string algebra \bref{NH}  is equivalent to the GCA \bref{GCA}.
So the GCA on the boundary is realized as the Newton-Hooke string algebra
in the bulk. 
The limit corresponds to taking a close-up 
of the \ads{2} string worldsheet.
Thus the GCA can be interpreted as the symmetry of the non-relativistic
string in \ads{}
space.

\medskip

The Newton-Hooke brane algebra
\cite{BGK} is a generalization of the Newton-Hooke string algebra to the case that
$\bar a$ spans a $p$-dimensional brane worldvolume.
Let $\bar a$ be $\bar a=(\alpha,d+1)$,
where $\alpha$ spans a $(p-1)$-dimensional spacetime.
The Newton-Hooke string algebra
is the case with $p=2$.
The scaling \bref{NH:scaling} means
\begin{eqnarray}
&&
\tilde P_\alpha=P_\alpha~,~~
\tilde K_\alpha=K_\alpha~,~~
\tilde D=D~,~~
\tilde J_{ij}=J_{ij}~,~~
\tilde J_{\alpha\beta}= J_{\alpha\beta}~,
\nonumber\\&&
\tilde P_i = \omega P_i~,~~
\tilde K_i = \omega  K_i~,~~
\tilde J_{i\alpha} = \omega  G_{i\alpha}~.
\end{eqnarray}
The Newton-Hooke brane algebra
is expressed on the boundary as
\begin{eqnarray}
[D,P_\alpha]&=&P_\alpha~,~~
[D,K_\alpha]=-K_\alpha~,~~
[D,P_i]=P_i~,~~
[D,K_i]=-K_i~,~~
\nonumber \\ 
{[P_\alpha,K_\beta]}&=&\frac{1}{2}J_{\alpha\beta}+\frac{1}{2}\eta_{\alpha\beta}D~,~~~
[P_i,K_\alpha]=\frac{1}{2}G_{i\alpha}~,~~~
[P_\alpha,K_i]=-\frac{1}{2}G_{i\alpha}~,~~~
\nonumber \\
{[J_{\alpha\beta}, P_{\gamma}]}&=&
\eta_{\beta\gamma} P_{\alpha}-\eta_{\alpha\gamma} P_{\beta}~,~~~
[J_{\alpha\beta}, K_{\gamma}]=
\eta_{\beta\gamma} K_{\alpha}-\eta_{\alpha\gamma} K_{\beta}~,~~~
[J_{\alpha\beta}, G_{i\gamma}]=
\eta_{\alpha\gamma} G_{\beta i}-\eta_{\beta\gamma} G_{\alpha i}~,~~~
\nonumber \\
{[J_{ij}, P_{k}]}&=&
\delta_{jk} P_{i}-\delta_{ik} P_{j}~,~~~
[J_{ij}, K_{k}]=
\delta_{jk} K_{i}-\eta_{ik} K_{j}~,~~~
[J_{ij}, G_{k\alpha}]=
\delta_{jk} G_{i\alpha}-\delta_{ik} G_{j\alpha}~,~~~
\nonumber \\
{[J_{ij}, J_{kl}]}&=&\delta_{jk} J_{il}+\mbox{3-terms}~,
~~~
[J_{\alpha\beta}, J_{\gamma\delta}]=\delta_{\beta\gamma} J_{\alpha\delta}+\mbox{3-terms}~.
\label{semi-GCA}
\end{eqnarray}
This is called as semi-GCA in \cite{Alishahiha:2009np}.
It reduces to the GCA \bref{GCA} when $\alpha=0$.

\medskip

In the subsequent sections,
we will derive super GCAs from 
superconformal algebras.

\section{Super GCA from
four-dimensional 
superconformal algebra
}

In this section, we derive super
GCAs
from
the four-dimensional
superconformal algebras, 
psu($2,2|4$) and su($2,2|2$).
\medskip

The bosonic part of psu($2,2|4$) is
the four-dimensional conformal algebra so($2,4$) given in \bref{conformal} with $\mu=0,1,2,3$,
and the R-symmetry so(6) $\cong $ su(4)
\begin{eqnarray}
[ \tilde P_{a'},  \tilde  P_{b'}]= - \tilde J_{a'b'}~,~~
[ \tilde J_{a'b'}, \tilde J_{c'd'}]=\delta_{b'c'} \tilde J_{a'd'}+\mbox{3-terms}~,~~
[ \tilde J_{a'b'},  \tilde P_{c'}]=\delta_{b'c'} \tilde P_{a'}-\delta_{a'c'}\tilde P_{b'}~,~~
\label{so}
\end{eqnarray}
where $a',b'=5,\cdots,9$.
The fermionic part of the (anti-)commutation relation is
\footnote{We follow the notation given in \cite{SY:Sch}.}
\begin{eqnarray}
[\tilde P_\mu, \CQ]&=&-\frac{1}{2}\CQ \Gamma_{\mu 4}p_+~,~~
[\tilde K_\mu, \CQ]=+\frac{1}{2}\CQ \Gamma_{\mu 4}p_-~,~~
[\tilde D, \CQ]=-\frac{1}{2}\CQ \Gamma_{4}
\CI i\sigma_2
~,~~
\nonumber\\
{[\tilde J_{\mu\nu},\CQ]}&=&\frac{1}{2}\CQ \Gamma_{\mu\nu}~,~~
[\tilde P_{a'},\CQ]=\frac{1}{2}\CQ \Gamma_{a'}\CJ i\sigma_2~,~~
[\tilde J_{a'b'},\CQ]=\frac{1}{2}\CQ \Gamma_{a'b'}~,~~
\nonumber\\
\{\CQ^T,\CQ\}&=&4iC\Gamma^\mu h_+(
 p_-\tilde P_\mu
+
p_+\tilde K_\mu
)
+2iC\Gamma^4 h_+ \tilde D
+iC\Gamma^{\mu\nu}\CI i\sigma_2 h_+ J_{\mu\nu}
\nonumber\\&&
+2iC\Gamma^{a'}h_+ \tilde  P_{a'}
-iC\Gamma^{a'b'}\CJ  i\sigma_2 h_+\tilde J_{a'b'}~,
\label{psu(2,2|4) fermionic}
\end{eqnarray}
where $\CI=\Gamma^{01234}$, $\CJ=\Gamma^{56789}$ and
\begin{eqnarray}
p_\pm=\frac{1}{2}(1\pm \Gamma^{0123}i\sigma_2)~.
\end{eqnarray}
The supercharge $\CQ=(Q_1,Q_2)$ is a pair of
16 component
 Majorana-Weyl spinors in ($1+9$)-dimensions
with the same chirality $Q_{1,2}=Q_{1,2}h_+$,
where $h_+$ is the chirality projector.
The $\tilde Q=\CQ p_-$ and $\tilde S=\CQ p_+$ correspond to
the supercharge
and the superconformal charge
of the $\CN=4$ superconformal algebra, respectively.

\subsection{32 supersymmetric GCA from psu($2,2|4$)}
We introduce a pair of  projectors defined by
\begin{eqnarray}
\ell_\pm=\frac{1}{2}(1\pm \Gamma^{04}\sigma)
\label{ell 4-dim}
\end{eqnarray}
where $\sigma=\sigma_1,\sigma_3$.
Note that 
$\ell_\pm$ commute with $h_+$, $p_+$ and $p_-$.
Decomposing $\CQ$ by using $\ell_\pm$ as
\begin{eqnarray}
\tilde Q_\pm = \CQ \ell_\pm~,
\end{eqnarray}
we rewrite \bref{psu(2,2|4) fermionic}
\begin{eqnarray}
[\tilde P_0,\tilde Q_\pm]&=&-\frac{1}{2}\tilde Q_\pm \Gamma_{04}p_+~,~~
[\tilde K_0,\tilde Q_\pm]=+\frac{1}{2}\tilde Q_\pm \Gamma_{04}p_-~,~~
[\tilde D, \tilde Q_\pm]= -\frac{1}{2}\tilde Q_\pm \Gamma_{4}\CI i\sigma_2~,~~
\nonumber\\
{[\tilde P_i,\tilde Q_\pm]}&=&-\frac{1}{2}\tilde Q_\mp \Gamma_{i4}p_+~,~~
[\tilde K_i,\tilde Q_\pm]=+\frac{1}{2}\tilde Q_\mp \Gamma_{i4}p_-~,~~
[\tilde J_{i0},\tilde Q_\pm]=\frac{1}{2}\tilde Q_\mp \Gamma_{i0}~,~~
\nonumber\\
{[\tilde J_{ij},\tilde Q_\pm]}&=&\frac{1}{2}\tilde Q_\pm \Gamma_{ij}~,~~
[\tilde P_{a'},\tilde Q_\pm]=\frac{1}{2}\tilde  Q_\mp \Gamma_{a'}\CJ i\sigma_2~,~~
[\tilde J_{a'b'},\tilde Q_\pm]=\frac{1}{2}\tilde Q_\pm \Gamma_{a'b'}~,~~
\nonumber\\
\{\tilde Q_\pm^T,\tilde Q_\pm\}&=&
4iC\Gamma^0h_+ \ell_\pm (p_- \tilde P_0
+
p_+
\tilde K_0)
+2iC\Gamma^4 h_+ \ell_\pm \tilde D
\nonumber\\&&
+iC\Gamma^{ij}\CI i\sigma_2 h_+\ell_\pm \tilde J_{ij}
-iC\Gamma^{a'b'}\CJ i\sigma_2 h_+\ell_\pm\tilde J_{a'b'}~,
\nonumber\\
\{\tilde Q_+^T,\tilde Q_-\}&=&
4iC\Gamma^i h_+ \ell_-(p_- \tilde P_i
+
p_+
\tilde K_i)
+2iC\Gamma^{i0}\CI i\sigma_2 h_+\ell_- \tilde J_{i0}
\nonumber\\&&
+2iC\Gamma^{a'} h_+ \ell_- \tilde  P_{a'}~,
\end{eqnarray}
where $i=1,2,3$.
$\tilde Q_\pm$ are a pair of 16 component spinors
which are independent of each other.

In addition to \bref{contraction}, we scale generators as
\begin{eqnarray}
\tilde Q_+= Q_+~,~~~
\tilde Q_-= \omega Q_-~,~~~
\tilde  P_{a'}=\omega   P_{a'}~,~~~
\tilde J_{a'b'}= J_{a'b'}~.
\end{eqnarray}
Substituting these into the above (anti-)commutation relation
and taking the limit $\omega\to \infty$,
we obtain
\begin{eqnarray}
[H, Q_\pm]&=&-\frac{1}{2}Q_\pm \Gamma_{04}p_+~,~~
[K, Q_\pm]=+\frac{1}{2}Q_\pm \Gamma_{04}p_-~,~~
[D, Q_\pm]= -\frac{1}{2}Q_\pm \Gamma_{4}\CI i\sigma_2~,~~
\nonumber\\
{[P_i,Q_+]}&=&-\frac{1}{2}Q_- \Gamma_{i4}p_+~,~~
[K_i,Q_+]=+\frac{1}{2}Q_- \Gamma_{i4}p_-~,~~
[G_i,Q_+]=\frac{1}{2}Q_- \Gamma_{i0}~,~~
\nonumber\\
{}[J_{ij},Q_\pm]&=&\frac{1}{2}Q_\pm \Gamma_{ij}~,~~
\label{GCA D3 fermi}
\end{eqnarray}
and
\begin{eqnarray}
{[J_{a'b'}, J_{c'd'}]}&=&\delta_{b'c'}   J_{a'd'}+\mbox{3-terms}~,~~~
[J_{a'b'}, P_{c'}]=\delta_{b'c'}   P_{a'}-\delta_{a'c'}  P_{b'}~,
\nonumber\\
{}[P_{a'},Q_+]&=&\frac{1}{2}Q_- \Gamma_{a'}\CJ i\sigma_2~,~~
[J_{a'b'}, Q_\pm]=\frac{1}{2}Q_\pm \Gamma_{a'b'}~,~~
\nonumber\\
\{Q_+^T, Q_+\}&=&
4iC\Gamma^0 h_+ \ell_+ (p_- H
+
p_+ 
K)
+2iC\Gamma^4 h_+ \ell_+ D
+iC\Gamma^{ij}\CI i\sigma_2 h_+\ell_+ J_{ij}
\nonumber\\&&
-iC\Gamma^{a'b'}\CJ i\sigma_2 h_+\ell_+ J_{a'b'}~,
\nonumber\\
\{ Q_+^T,Q_-\}&=&
4iC\Gamma^i h_+ \ell_-  ( p_-P_i
+
p_+ 
K_i)
+2iC\Gamma^{i0}\CI i\sigma_2 h_+\ell_- G_i
\nonumber\\&&
+2iC\Gamma^{a'} h_+ \ell_-  P_{a'}~,
\end{eqnarray}
in addition to \bref{GCA}.
This is a 32 supersymmetric GCA.
The bosonic subalgebra is the GCA and iso($5$).
By construction, the super GCA
obtained above
\begin{eqnarray}
\{H,~
K,~
D,~
P_i,~K_i,~G_i,~
J_{ij},~
 P_{a'},~
 J_{a'b'},~
Q_+,~Q_-
\}
\end{eqnarray}
coincides with the super Newton-Hooke algebra
of a string in \adss{5}{5}
\cite{GGK}\cite{SY:NH}.
The string lies along the 4-th direction which is transverse to the boundary of \ads{5}
and the worldsheet is \ads{2}.
Thus the super GCA obtained above is a boundary realization of the super Newton-Hooke
algebra of the \ads{2} string.

We note that the set of generators
\begin{eqnarray}
\{H,~K,~D,~J_{ij},~J_{a'b'},~
Q_+\}
\label{subalgebra D3}
\end{eqnarray}
forms a 16 supersymmetric subalgebra.
$\{H,K,D\}$ corresponds  to the symmetry of the \ads{2}.
$J_{ij}$ and $J_{a'b'}$ are the rotational symmetry 
in the space transverse to \ads{2} in \adss{5}{5}.
The subalgebra \bref{subalgebra D3}
is the residual symmetry  in the presence of a 1/2 BPS string.
It is seen from the superalgebra
that the string is F(D)-string for $\sigma=\sigma_3(\sigma_1$, respectively).

\subsection{16 supersymmetric GCA from su($2,2|2$)}

Firstly, we derive the $\CN=2$ superconformal algebra su($2,2|2$)
from
the $\CN=4$ superconformal algebra psu($2,2|4$).
For this, we introduce a projector \cite{SY:Sch_less}
\begin{eqnarray}
q_+=\frac{1}{2}(1+\Gamma^{5678})~,
\end{eqnarray}
and require that
\begin{eqnarray}
\CQ\equiv \CQ q_+~.
\end{eqnarray}
We note that $q_+$ commutes with  $h_+$ and $p_\pm$.
The fermionic part of
psu($2,2|4$)
reduces to
\begin{eqnarray}
[\tilde P_\mu, \CQ]&=&-\frac{1}{2}\CQ \Gamma_{\mu 4}p_+~,~~
[\tilde K_\mu, \CQ]=+\frac{1}{2}\CQ \Gamma_{\mu 4}p_-~,~~
[\tilde D, \CQ]=-\frac{1}{2}\CQ \Gamma_{4}
\CI i\sigma_2
~,~~
\nonumber\\
{}[\tilde J_{\mu\nu},\CQ]&=&\frac{1}{2}\CQ \Gamma_{\mu\nu}~,~~
[\tilde P_{9},\CQ]=\frac{1}{2}\CQ \Gamma_{9}\CJ i\sigma_2~,~~
[\tilde J_{mn},\CQ]=\frac{1}{2}\CQ \Gamma_{mn}~,~~
\nonumber\\
\{\CQ^T,\CQ\}&=&4iC\Gamma^\mu h_+q_+(p_-\tilde P_\mu
+
p_+
\tilde K_\mu)
+2iC\Gamma^4 h_+q_+ \tilde D
+iC\Gamma^{\mu\nu}\CI i\sigma_2 h_+q_+\tilde J_{\mu\nu}
\nonumber\\&&
+2iC\Gamma^{9}h_+q_+ \tilde P_{9}
-iC\Gamma^{mn}\CJ  i\sigma_2 h_+q_+\tilde J_{mn}~,
\label{N=2 fermionic}
\end{eqnarray}
where $m=5,6,7,8$.
The bosonic part 
is \bref{conformal}
and
su($2)^2\times$u(1)
generated by 
$\{\tilde P_9, \tilde J_{mn}\}$
\begin{eqnarray}
[\tilde J_{mn},\tilde J_{pq}]=\delta_{np}\tilde J_{mq}+\mbox{3-terms}~.
\end{eqnarray}
Only the su($2)\times$u(1) part in su($2)^2\times$u(1)
acts non-trivially on the supercharge $\CQ$.
To see this, we note that
the commutation relations containing $\tilde J_{mn}$
can be rewritten as
\begin{eqnarray}
&&
[\tilde J_I^{(\pm)},\tilde J_J^{(\pm)}]=\epsilon_{IJK}\tilde J_K^{(\pm)}~,~~
[\tilde J_I^{(+)},\CQ]=\frac{1}{2}\CQ \rho_I~,~~
[\tilde J_I^{(-)},\CQ]=0~,
\nonumber\\&&
\{\CQ^T,\CQ\}=\cdots
-4iC\CJ i\sigma_2 h_+q_+ 
\left(
\rho_1\tilde J_1^{(+)}+\rho_2\tilde J_2^{(+)}+\rho_3\tilde J_3^{(+)}
\right)~,
\end{eqnarray}
which follow from the fact $\Gamma^{5678}q_+=q_+$.
We have defined
\begin{eqnarray}
\tilde J_I^{(\pm)}=
\left(
\frac{1}{2}(\tilde J_{56}\mp \tilde J_{78})\,,~
\frac{1}{2}(\tilde J_{57}\pm \tilde J_{68})\,,~
\frac{1}{2}(\pm \tilde J_{58}-\tilde J_{67})
\right)~,~~
\rho_I\equiv (\Gamma^{56},\Gamma^{57},\Gamma^{58})~,
\end{eqnarray}
and  $\epsilon_{123}=1$.
The $\CN=2$ superconformal algebra,
\begin{eqnarray}
\{\tilde P_\mu,~
\tilde K_\mu,~
\tilde D,~
\tilde J_{\mu\nu},~
\tilde P_9,~
J_I^{(+)},~
\CQ
\}~,
\end{eqnarray}
is su($2,2|2$).
The set of generators
$\{J_I^{(+)},\tilde P_9\}$ acts as the R-symmetry su(2)$\times$u(1).

\medskip
Next,
we introduce $\ell_\pm$ in \bref{ell 4-dim},
which commute with $q_+$,
and decompose $\CQ$ as $\tilde Q_\pm=\CQ\ell_\pm$.
In addition to \bref{contraction}, we scale generators as
\begin{eqnarray}
\tilde Q_+=Q_+~,~~\tilde Q_-=\omega Q_-~,~~
\tilde J_I^{(+)}=J_I^{(+)}~,~~
\tilde P_9=\omega P_9~.
\end{eqnarray}
Under the contraction, we obtain a 16 supersymmetric GCA from su($2,2|2$)
\begin{eqnarray}
\{H,~
K,~
D,~
J_{ij},~
P_i,~
K_i,~
G_i,~
P_9,~
J_I^{(+)},~
Q_+,~
Q_-\}~,
\end{eqnarray}
with the 
(anti-)commutation relations, 
\bref{GCA}, \bref{GCA D3 fermi} and 
\begin{eqnarray}
[J_I^{(+)},J_J^{(+)}]&=&\epsilon_{IJK}J_K^{(+)}~,~~
[P_{9},Q_+]=\frac{1}{2}Q_- \Gamma_{9}\CJ i\sigma_2~,~~
[J_I^{(+)},Q_\pm]=\frac{1}{2}Q_\pm \rho_I~,~~
\nonumber\\
\{Q_+^T, Q_+\}&=&
4iC\Gamma^0 h_+ \ell_+q_+(p_-  H
+
p_+ K)
+2iC\Gamma^4 h_+ \ell_+q_+ D
+iC\Gamma^{ij}\CI i\sigma_2 h_+\ell_+q_+ J_{ij}
\nonumber\\&&
-4iC\CJ i\sigma_2 h_+q_+\ell_+ 
\left(
\rho_1J_1^{(+)}+\rho_2J_2^{(+)}+\rho_3J_3^{(+)}
\right)
~,
\nonumber\\
\{ Q_+^T,Q_-\}&=&
4iC\Gamma^i h_+ \ell_-q_+ (p_-  P_i
+
p_+ K_i)
+2iC\Gamma^{i0}\CI i\sigma_2 h_+\ell_-q_+ G_i
\nonumber\\&&
+2iC\Gamma^{9} h_+ \ell_-q_+  P_{9}~.
\label{GCA D3 N=2}
\end{eqnarray}
We note that the set of generators
\begin{eqnarray}
\{H,~K,~D,~J_{ij},~J_I^{(+)},~
Q_+\}
\label{subalgebra D3 N=2}
\end{eqnarray}
forms a 8 supersymmetric subalgebra.
$J_I^{(+)}$ acts as the R-symmetry su(2).
\medskip

Introducing the 1/4 projector as in  \cite{SY:Sch_less},
$q_+=\frac{1}{2}(1+\Gamma^{56}i\sigma_2)\frac{1}{2}(1+\Gamma^{78}i\sigma_2)$,
we can derive the $\CN=1$ superconformal algebra
su($2,2|1$)
 by considering the diagonal u(1) in  u(1)$^3$.
But the $\ell_\pm$ does not commute with $q_+$.
This is because $\CN=1$ is the minimal supersymmetry
and to derive our super GCA we need at least two sets of supercharges,
namely $\CN=2$.

\section{Super GCA from
three-dimensional superconformal algebra
}
In this section, we derive super GCAs from the three-dimensional superconformal 
algebras osp($8|4$), osp($4|4$) and  osp($2|4$).
\medskip

The super-\adss{4}{7} algebra osp($8|4$)
is the $\CN=8$ superconformal algebra in three-dimensions.
The bosonic part of the algebra is 
given in \bref{conformal} with $\mu=0,1,2$
and \bref{so} with $a'=4,\cdots,9,\natural$.
The fermionic part is
\begin{eqnarray}
[\tilde P_\mu,\CQ]&=&-\frac{1}{2}\CQ \Gamma_{\mu 3}p_+~,
~~~
[\tilde K_\mu,\CQ]=+\frac{1}{2}\CQ \Gamma_{\mu 3}p_-~,
~~~
[\tilde D,\CQ]=-\frac{1}{2}\CQ\CI  \Gamma_{3}~,
~~~
[\tilde J_{\mu\nu},\CQ]=\frac{1}{2}\CQ \Gamma_{\mu \nu}~,
\nonumber\\
{}[\tilde P_{a'},\CQ]&=&-\frac{1}{2}\CQ\CI \Gamma_{a'}~,
~~~
[\tilde J_{a'b'},\CQ]=\frac{1}{2}\CQ  \Gamma_{a'b'}~,
\nonumber\\
\{\CQ^T,\CQ\}&=&
-4C\Gamma^\mu(p_+\tilde K_\mu +p_-\tilde P_\mu )
-2C\Gamma^3 \tilde D
+C\CI \Gamma^{\mu\nu}\tilde J_{\mu\nu}
-C\Gamma^{a'}\tilde P_{a'}
-\frac{1}{2}C\CI\Gamma^{a'b'}\tilde J_{a'b'}
~,
\end{eqnarray}
where $\CI=\Gamma^{0123}$ and 
\begin{eqnarray}
p_\pm=\frac{1}{2}(1\pm\Gamma^{012})~.
\end{eqnarray}
The supercharge $\CQ$ is a 32 component Majorana spinor in $(1+10)$-dimensions.
$\{\tilde P_{a'},\tilde J_{a'b'}\}$ generates the R-symmetry so($8$).

\subsection{32 supersymmetric GCA from osp($8|4$)}

Let us decompose $\CQ$ as $\tilde Q_\pm=\CQ\ell_\pm$
by introducing a pair of projectors
\begin{eqnarray}
\ell_\pm=\frac{1}{2}(1\pm \Gamma^{034})~,
\label{ell M2}
\end{eqnarray}
which commute with $p_\pm$,
as $[\Gamma^{034},p_\pm]=0$.
In addition to \bref{contraction}, we scale generators as
\begin{eqnarray}
\tilde Q_+= Q_+~,~~
\tilde Q_-= \omega Q_-~,~~
\tilde P_{4}=P_{4}~,~~
\tilde P_{m}=\omega   P_{m}~,~~
\tilde J_{4m}= \omega J_{4m}~,~~
\tilde J_{mn}= J_{mn}~,
\end{eqnarray}
where $a'=(4,m)$ and $m=5,6,7,8,9,\natural$.
Substituting these into the (anti-)commutation relation
of the $\CN=8$ superconformal algebra
and taking the limit $\omega\to \infty$,
we arrive at
\begin{eqnarray}
[H, Q_\pm]&=&-\frac{1}{2}Q_\pm \Gamma_{03}p_+~,~~
[K, Q_\pm]=+\frac{1}{2}Q_\pm \Gamma_{03}p_-~,~~
[D, Q_\pm]= -\frac{1}{2}Q_\pm \CI \Gamma_{3}~,~~
\nonumber\\
{}[P_i,Q_+]&=&-\frac{1}{2}Q_- \Gamma_{i3}p_+~,~~
[K_i,Q_+]=+\frac{1}{2}Q_- \Gamma_{i3}p_-~,~~
[G_i,Q_+]=\frac{1}{2}Q_- \Gamma_{i0}~,~~
\nonumber\\
{}[J_{ij},Q_\pm]&=&\frac{1}{2}Q_\pm \Gamma_{ij}~,~~
\label{GCA M2 fermi}
\end{eqnarray}
and
\begin{eqnarray}
[J_{mn}, J_{pq}]&=&\delta_{np}   J_{mq}+\mbox{3-terms}~,~~~
[J_{mn}, P_{p}]=\delta_{np}   P_{m}-\delta_{mp}  P_{n}~,
\nonumber\\
{}[J_{mn}, J_{4p}]&=&\delta_{mp}   J_{n4}-\delta_{np}  J_{m4}~,~~~
[J_{4m}, P_{4}]=-P_{m}~,
\nonumber\\
{}[P_{4},Q_\pm]&=&-\frac{1}{2}Q_\pm\CI \Gamma_{4}~,~~
[P_{m},Q_+]=-\frac{1}{2}Q_- \CI\Gamma_{m}~,~~
\nonumber\\
{}[J_{mn}, Q_\pm]&=&\frac{1}{2}Q_\pm \Gamma_{mn}~,~~
[J_{4m}, Q_+]=\frac{1}{2}Q_- \Gamma_{4m}~,~~
\nonumber\\
\{Q_+^T, Q_+\}&=&
-4C\Gamma^0\ell_+(p_- H
+p_+ K)
-2C\Gamma^3 \ell_+ D
+C\CI\Gamma^{ij}\ell_+ J_{ij}
\nonumber\\&&
-C\Gamma^{4}\ell_+ P_{4}
-\frac{1}{2}C\CI\Gamma^{mn}\ell_+ J_{mn}~,
\nonumber\\
\{ Q_+^T,Q_-\}&=&
-4C\Gamma^i\ell_- ( p_-  P_i
+p_+   K_i)
+2C\CI\Gamma^{i0}\ell_- G_i
-C\Gamma^{m}\ell_-  P_{m}
-C\Gamma^{4m}\ell_-  J_{4m}
~,
\end{eqnarray}
in addition to \bref{GCA} with $i=1,2$.
This is a 32 supersymmetric GCA.
By construction, the super GCA
\begin{eqnarray}
\{H,~
K,~
D,~
P_i,~K_i,~G_i,~
J_{ij},~
 P_{4},~
P_m,~
J_{mn},~
J_{4m},~
Q_+,~Q_-
\}
\end{eqnarray}
is the super Newton-Hooke algebra of an M2-brane in \adss{4}{7}.
The M2-brane worldvolume is \adss{2}{1} 
extending along $(0,3,4)$-th directions.
The M2-brane worldvolume extends along two directions in \ads{4} 
and so the GCA results.
If a brane extends along more than two directions in \ads{},
a semi-GCA emerges generally as seen in the next section.
The super GCA obtained above contains a 16 supersymmetric subalgebra
\begin{eqnarray}
\{H,~
K,~
D,~
J_{ij},~
P_{4},~
J_{mn},~
Q_+
\}~.
\end{eqnarray}
The set of generators
$\{H,K,D,P_4\}$
generates the \adss{2}{1} symmetry on the M2-brane worldvolume,
while $\{J_{ij},J_{mn}\}$ is the rotational symmetry
so($2$)$\times$so(6) in the space
 transverse to  \adss{2}{1} in \adss{4}{7}.

\subsection{16 supersymmetric GCA from osp($4|4$)}
We derive a 16 supersymmetric GCA from 
the $\CN=4$ superconformal algebra osp($4|4$).

Firstly, 
we derive  osp($4|4$) from
the $\CN=8$ superconformal algebra osp($8|4$).
For this, we introduce a 1/2 projector in the $\CN=8$ superconformal algebra
\begin{eqnarray}
q_+=\frac{1}{2}(1+\Gamma^{789\natural})~,
\end{eqnarray}
and require that $\CQ\equiv \CQ q_+$.
We note $q_+$ commutes with $p_\pm$.
Then the $\CN=8$ superconformal algebra reduces to
a $\CN=4$ superconformal algebra.
The fermionic part of the algebra is
\begin{eqnarray}
[\tilde P_\mu,\CQ]&=&-\frac{1}{2}\CQ \Gamma_{\mu 3}p_+~,
~~~
[\tilde K_\mu,\CQ]=+\frac{1}{2}\CQ \Gamma_{\mu 3}p_-~,
~~~
[\tilde D,\CQ]=-\frac{1}{2}\CQ\CI  \Gamma_{3}~,
~~~
[\tilde J_{\mu\nu},\CQ]=\frac{1}{2}\CQ \Gamma_{\mu \nu}~,
\nonumber\\
{}[P_m,\CQ]&=&-\frac{1}{2}\CQ\CI \Gamma_m~,~~~
[J_{mn},\CQ]=\frac{1}{2}\CQ \Gamma_{mn}~,~~~
[J_{m'n'},\CQ]=\frac{1}{2}\CQ \Gamma_{m'n'}~,
\nonumber\\
\{\CQ^T,\CQ\}&=&
-4C\Gamma^\mu q_+( p_+\tilde K_\mu+ p_-\tilde P_\mu)
-2C\Gamma^3q_+ \tilde D
+C\CI \Gamma^{\mu\nu}q_+\tilde J_{\mu\nu}
\nonumber\\&&
-C\Gamma^{m}q_+\tilde P_{m}
-\frac{1}{2}C\CI q_+(\Gamma^{mn}\tilde J_{mn}+\Gamma^{m'n'}\tilde J_{m'n'})
~,
\end{eqnarray}
where $a'=(m,m')$, $m=4,5,6$ and $m'=7,8,9,\natural$.
The bosonic part is \bref{conformal} and 
 so(4)$\times$so(4)
generated by
$\{\tilde P_m,\tilde J_{mn}\}$
and $\{J_{m'n'}\}$.

As was done in section 3.2,
we rewrite the commutation relations containing $J_{m'n'}$
as
\begin{eqnarray}
[\tilde J_I'^{(\pm)},\tilde J_I'^{(\pm)}]&=&\epsilon_{IJK}\tilde J_K'^{(\pm)}~,~~
[\tilde J_I'^{(+)},\CQ]=\frac{1}{2}\CQ \rho_I'~,~~
[\tilde J_I'^{(-)},\CQ]=0~,
\nonumber\\
\{\CQ^T,\CQ\}&=&\cdots -2C\CI q_+
\sum_{I=1,2,3}
\rho_I'\tilde J_I'^{(+)}
\end{eqnarray}
with
\begin{eqnarray}
\tilde J_I'^{(\pm)}=\left(
\frac{1}{2}(\tilde J_{89}\mp \tilde J_{7\natural}),~
\frac{1}{2}(\tilde J_{79}\pm \tilde J_{8\natural}),~
\frac{1}{2}(\tilde J_{78}\mp \tilde J_{9\natural})
\right)~,~~
\rho_I'=(\Gamma_{89},\Gamma_{79},\Gamma_{78})~.
\end{eqnarray}
Similarly,
the commutation relations containing $\tilde P_m $ and $\tilde J_{mn}$
can be rewritten as
\begin{eqnarray}
[\tilde J_I^{(\pm)},\tilde J_I^{(\pm)}]&=&\epsilon_{IJK}\tilde J_K^{(\pm)}~,~~
[\tilde J_I^{(+)},\CQ]=\frac{1}{2}\CQ \rho_I~,~~
[\tilde J_I^{(-)},\CQ]=0~,
\nonumber\\
\{\CQ^T,\CQ\}&=&\cdots -2C\CI q_+
\sum_{I=1,2,3}
\rho_I\tilde J_I^{(+)}
\end{eqnarray}
with
\begin{eqnarray}
\tilde J_I^{(\pm)}=\left(
\frac{1}{2}(\tilde P_6\pm \tilde J_{45}),~
\frac{1}{2}(\tilde P_5 \mp \tilde J_{46}),~
\frac{1}{2}(\pm\tilde  P_4 +\tilde  J_{56})
\right)~,~~
\rho_I=(\Gamma_{45},-\Gamma_{46},\Gamma_{56})~.
\end{eqnarray}
These follow from the fact $\Gamma^{789\natural}q_+=q_+$ 
and $\Gamma^\natural=\Gamma^{0\cdots 9}$.
Thus the set of generators
\begin{eqnarray}
\{\tilde P_\mu,~
\tilde K_\mu,~
\tilde D,~
\tilde J_{\mu\nu},~
\tilde J_I^{(+)},~
\tilde J_I'^{(+)},~
\CQ
\}
\end{eqnarray}
forms a superconformal algebra, which is osp($4|4$).
$\{\tilde J_I^{(+)},\tilde J_I'^{(+)}\}$ 
generates the R-symmetry 
su(2)$\times$su(2)
$\cong$
so(4).

\medskip

Next, 
we decompose $\CQ$ as $\tilde Q_\pm=\CQ\ell_\pm$
by introducing 
a pair of projectors $\ell_\pm$ in \bref{ell M2},
which commute with $q_+$,
and scale
generators as \bref{contraction} and
\begin{eqnarray}
&&
\tilde Q_+= Q_+~,~~
\tilde Q_-= \omega Q_-~,~~
\tilde J_3^{(+)}= J_3^{(+)}~,~~
\tilde J_{1,2}^{(+)}=\omega J_{1,2}^{(+)}~,~~
\tilde J_I'^{(+)}= J_I'^{(+)}~.
\label{scaling M2 N=4}
\end{eqnarray}
Substituting these into the (anti-)commutation relation
of the $\CN=4$ superconformal algebra osp($4|4$)
and taking the limit $\omega\to \infty$,
we derive \bref{GCA}, \bref{GCA M2 fermi} and
\begin{eqnarray}
[J_3^{(+)},J_I^{(+)}]&=&\epsilon_{3IJ}J_J^{(+)}~,~~
[J_3^{(+)},Q_\pm]=-\frac{1}{2}Q_\pm\CI \rho_3~,~~
[J_{1,2}^{(+)},Q_+]=-\frac{1}{2}Q_- \CI\rho_{1,2}~,~~
\nonumber\\
{}[J_I'^{(+)},J_J'^{(+)}]&=&\epsilon_{IJK}J_K'^{(+)}~,~~
[J_I'^{(+)},Q_\pm]=\frac{1}{2}Q_\pm \rho_I'~,~~
\nonumber\\
\{Q_+^T, Q_+\}&=&
-4C\Gamma^0\ell_+q_+(p_- H
+p_+  K)
-2C\Gamma^3 \ell_+q_+ D
+C\CI\Gamma^{ij}\ell_+q_+ J_{ij}
\nonumber\\&&
-2C\rho_3\ell_+q_+ J_3^{(+)}
-2C\CI \ell_+q_+ \sum_{I=1,2,3}
\rho_I'J_I'^{(+)} ,
\nonumber\\
\{ Q_+^T,Q_-\}&=&
-4C\Gamma^i \ell_-q_+ ( p_- P_i
+p_+ K_i)
+2C\CI\Gamma^{i0}\ell_-q_+ G_i
\nonumber\\&&
-2C\CI q_+ (\rho_1J_1^{(+)}+\rho_2J_2^{(+)})\ell_-
~.
\label{GCA M2 N=4}
\end{eqnarray}
This is a 16 supersymmetric GCA
\begin{eqnarray}
\{H,~
K,~
D,~
J_{ij},~
P_i,~
K_i,~
G_i,~
J_I^{(+)},~
J_I'^{(+)},~
Q_+,~
Q_-\}
~.
\end{eqnarray}
The bosonic subalgebra is the GCA and so(3)$\times$iso($2$).
The super GCA contains a 8 supersymmetric subalgebra
\begin{eqnarray}
\{H,~
K,~
D,~
J_{ij},~
J_3^{(+)},~
J_I'^{(+)},~
Q_+
\}~.
\end{eqnarray}
$\{J_I'^{(+)},J_3^{(+)}\}$ generates the R-symmetry su(2)$\times$u(1).

\subsection{8 supersymmetric GCA from osp($2|4$)}
We derive a 8 supersymmetric GCA from the $\CN=2$ superconformal algebra
osp($2|4$).

Firstly, we derive osp($2|4$) from the $\CN=8$ superconformal algebra osp($8|4$).
We introduce a 1/4 projector
\begin{eqnarray}
q_+=\frac{1}{2}(1+\Gamma^{5678})\frac{1}{2}(1+\Gamma^{789\natural})~,
\end{eqnarray}
which commutes with $p_\pm$,
and require that $\CQ\equiv \CQ q_+$.
Then the $\CN=8$ superconformal algebra reduces to
a $\CN=2$ superconformal algebra.
The fermionic part of the algebra is
\begin{eqnarray}
[P_4,\CQ]&=&-\frac{1}{2}\CQ\CI \Gamma_4~,~~~
[J_{56},\CQ]=\frac{1}{2}\CQ \Gamma_{56}~,~~~
[J_{78},\CQ]=\frac{1}{2}\CQ \Gamma_{78}~,~~~
[J_{9\natural},\CQ]=\frac{1}{2}\CQ \Gamma_{9\natural}~,~~~
\nonumber\\
\{\CQ^T,\CQ\}&=&
-4C\Gamma^\mu q_+( p_+\tilde K_\mu+p_-\tilde P_\mu )
-2C\Gamma^3q_+ \tilde D
+C\CI \Gamma^{\mu\nu}q_+\tilde J_{\mu\nu}
\nonumber\\&&
-C\Gamma^{4}q_+\tilde P_{4}
-C\CI q_+(\Gamma^{56}\tilde J_{56}+\Gamma^{78}\tilde J_{78}+\Gamma^{9\natural}\tilde J_{9\natural})
~.
\end{eqnarray}
The bosonic part is \bref{conformal} and
u(1)$^4$
generated by
$\{\tilde P_4,\tilde J_{56},\tilde J_{78},\tilde J_{9\natural}\}$.
We note 
the commutation relations containing u(1)$^4$ generators 
are rewritten as
\begin{eqnarray}
[\tilde J_I,\CQ]=-\frac{1}{2}\CQ\CI\Gamma_4~,~~
\{\CQ^T,\CQ\}=\cdots
-4C\Gamma^4q_+\tilde J
~,
\end{eqnarray}
with
\begin{eqnarray}
\tilde J_I=\left(
\tilde P_4,
\tilde J_{56},
 -\tilde J_{78},
\tilde J_{9\natural}
\right)~,~~~
\tilde J=\frac{1}{4}\sum_{I=1,2,3,4}\tilde J_I~.
\end{eqnarray}
It follows that the set of generators
\begin{eqnarray}
\{\tilde P_\mu,~
\tilde K_\mu,~
\tilde D,~
\tilde J_{\mu\nu},~
\tilde J,~
\CQ
\}
\end{eqnarray}
forms the $\CN=2$ superconformal algebra osp($2|4$).
$\tilde J$ generates the R-symmetry so(2).

\medskip

Next,
we decompose $\CQ$ as $\tilde Q_\pm = \CQ\ell_\pm$ by $\ell_\pm$ in \bref{ell M2},
which commute with $q_+$,
and scale
generators as \bref{contraction} and
\begin{eqnarray}
\tilde Q_+= Q_+~,~~~
\tilde Q_-= \omega Q_-~,~~~
\tilde J =J
~.
\end{eqnarray}
Substituting these into the above (anti-)commutation relation
of osp($2|4$)
and taking the limit $\omega\to \infty$,
we derive \bref{GCA}, \bref{GCA M2 fermi} and
\begin{eqnarray}
[J,Q_\pm]&=&-\frac{1}{2}Q_\pm\CI \Gamma_4~,~~
\nonumber\\
\{Q_+^T, Q_+\}&=&
-4C\Gamma^0\ell_+q_+(p_- H
+p_+ K)
-2C\Gamma^3 \ell_+q_+ D
+C\CI\Gamma^{ij}\ell_+q_+ J_{ij}
\nonumber\\&&
-4C\Gamma^4q_+\ell_+J
~,
\nonumber\\
\{ Q_+^T,Q_-\}&=&
-4C\Gamma^i\ell_-q_+( p_-   P_i
+p_+  K_i)
+2C\CI\Gamma^{i0}\ell_-q_+ G_i
~.
\label{GCA M2 N=2}
\end{eqnarray}
This is a 8 supersymmetric GCA
\begin{eqnarray}
\{H,~
K,~
D,~
J_{ij},~
P_i,~
K_i,~
G_i,~
J,~
Q_+,~
Q_-\}
~.
\end{eqnarray}
It contains a 4 supersymmetric subalgebra
\begin{eqnarray}
\{H,~
K,~
D,~
J_{ij},~
J,~
Q_+
\}
~,
\end{eqnarray}
in which $J$ acts as the R-symmetry so(2).

\section{Super semi-GCA from
six-dimensional superconformal algebra
}
In this section, we derive super GCAs from the
six-dimensional superconformal algebras,
osp  osp($8^*|4$) and osp($8^*|2$). 
\medskip

The super-\adss{7}{4} algebra osp($8^*|4$)  is
the $\CN=4$ superconformal algebra in six-dimensions.
The bosonic part is
\bref{conformal}
with $\mu=0,1,\cdots,5$
and the R-symmetry sp(4)\,$\cong$\,so(5)
given in \bref{so} with $a'=7,8,9,\natural$.
The fermionic part is 
\begin{eqnarray}
[\tilde P_\mu,\CQ]&=&-\frac{1}{2}\CQ \Gamma_{\mu 6}p_+~,
~~~
[\tilde K_\mu,\CQ]=+\frac{1}{2}\CQ \Gamma_{\mu 6}p_-~,
~~~
[\tilde D,\CQ]=-\frac{1}{2}\CQ\CI  \Gamma_{6}~,
~~~
[\tilde J_{\mu\nu},\CQ]=\frac{1}{2}\CQ \Gamma_{\mu \nu}~,
\nonumber\\
{}[\tilde P_{a'},\CQ]&=&-\frac{1}{2}\CQ\CI \Gamma_{a'}~,
~~~
[\tilde J_{a'b'},\CQ]=\frac{1}{2}\CQ  \Gamma_{a'b'}~,
\nonumber\\
\{\CQ^T,\CQ\}&=&
-4C\Gamma^\mu( p_+\tilde K_\mu+ p_-\tilde P_\mu)
-2C\Gamma^6 \tilde D
-C\CI \Gamma^{\mu\nu}\tilde J_{\mu\nu}
-4C\Gamma^{a'}\tilde P_{a'}
+2C\CI\Gamma^{a'b'}\tilde J_{a'b'}
~,
\label{superalgebra M5 N=4}
\end{eqnarray}
where $\CI=\Gamma^{789\natural}$
and
\begin{eqnarray}
p_\pm=\frac{1}{2}(1\pm \Gamma^{6789\natural})~.
\end{eqnarray}
The supercharge $\tilde Q=\CQ p_-$
(and  $\tilde S=\CQ p_+$ also) 
is composed of  four sets of four-component Weyl spinors
subject to the sp(4) Majorana condition.
The sp(4)\,$\cong$\,so(5) is generated by $\{P_{a'},J_{a'b'}\}$.

\subsection{32 supersymmetric semi-GCA from osp($8^*|4$)}
Let us decompose 
$\CQ$ as $\tilde Q_\pm=\CQ\ell_\pm$
by
introducing a pair of projectors
\begin{eqnarray}
\ell_\pm=\frac{1}{2}(1\pm \Gamma^{056})~,
\label{ell M5}
\end{eqnarray}
which commute with $p_\pm$,
and scale generators as
\begin{eqnarray}
&&
\tilde P_\alpha=P_\alpha~,~~~
\tilde K_\alpha=K_\alpha~,~~~
\tilde J_{05}=J_{05}~,~~~
\tilde J_{ij}=J_{ij}~,~~~
\tilde Q_+= Q_+~,~~~
\nonumber\\&&
\tilde P_{i}=\omega P_{i}~,~~~
\tilde K_{i}=\omega K_{i}~,~~~
\tilde J_{i\alpha}=\omega G_{i\alpha}~,~~~
\tilde Q_-= \omega Q_-~,~~~
\label{scaling M5}
\end{eqnarray}
and
\begin{eqnarray}
\tilde P_{a'}=\omega P_{a'}~,~~~
\tilde J_{a'b'}=J_{a'b'}~,~~~
\end{eqnarray}
where $\alpha=0,5$ and $i=1,2,3,4$.
Substituting these into the (anti-)commutation relation
of the $\CN=4$ superconformal algebra
osp($8^*|4$)
and taking the limit $\omega\to \infty$,
we derive a 32 supersymmetric semi-GCA.
The bosonic part is
\bref{semi-GCA}
and
\begin{eqnarray}
[J_{a'b'}, J_{c'd'}]&=&\delta_{b'c'}   J_{a'd'}+\mbox{3-terms}~,~~~
[J_{a'b'}, P_{c'}]=\delta_{b'c'}   P_{a'}-\delta_{a'c'}   P_{b'}~.
\end{eqnarray}
The fermionic part is
\begin{eqnarray}
[P_\alpha, Q_\pm]&=&-\frac{1}{2}Q_\pm \Gamma_{\alpha 6} p_+~,~~
[K_\alpha, Q_\pm]=+\frac{1}{2}Q_\pm \Gamma_{\alpha 6}p_-~,~~
[D, Q_\pm]= -\frac{1}{2}Q_\pm \CI \Gamma_{6}~,~~
\nonumber\\
{[P_i,Q_+]}&=&-\frac{1}{2}Q_- \Gamma_{i6}p_+~,~~
[K_i,Q_+]=+\frac{1}{2}Q_- \Gamma_{i6}p_-~,~~
[G_{i\alpha},Q_+]=\frac{1}{2}Q_- \Gamma_{i\alpha}~,~~
\nonumber\\
{[J_{ij},Q_\pm]}&=&\frac{1}{2}Q_\pm \Gamma_{ij}~,~~
[J_{\alpha\beta}, Q_\pm]=\frac{1}{2}Q_\pm \Gamma_{\alpha\beta}~,~~
\label{GCA M5 fermi}
\end{eqnarray}
and
\begin{eqnarray}
{[P_{a'},Q_+]}&=&-\frac{1}{2}Q_-\CI \Gamma_{a'}~,~~
[J_{a'b'},Q_\pm]=\frac{1}{2}Q_\pm \Gamma_{a'b'}~,~~
\nonumber\\
\{Q_+^T, Q_+\}&=&
-4C\Gamma^\alpha\ell_+(p_- P_\alpha
+p_+ K_\alpha)
-2C\Gamma^6 \ell_+ D
-C\CI\Gamma^{\alpha\beta}\ell_+ J_{\alpha\beta}
-C\CI\Gamma^{ij}\ell_+ J_{ij}
\nonumber\\&&
+2C\CI\Gamma^{a'b'}\ell_+ J_{a'b'}~,
\nonumber\\
\{ Q_+^T,Q_-\}&=&
-4C\Gamma^i\ell_-( p_-P_i
+p_+K_i)
-2C\CI\Gamma^{i\alpha}\ell_- G_{i\alpha}
-4C\Gamma^{a'}\ell_-  P_{a'}
~.
\end{eqnarray}
By construction, the supersymmetric  semi-GCA
\begin{eqnarray}
\{P_\alpha,~
K_\alpha,~
D,~
P_i,~K_i,~G_{i\alpha},~
J_{ij},~
J_{05},~
P_{a'},~
J_{a'b'},~
Q_+,~Q_-
\}
\end{eqnarray}
is the super Newton-Hooke algebra of an M2-brane in \adss{7}{4}.
The M2-brane worldvolume
is \ads{3} extending along $(0,5,6)$-th directions in \ads{7}.
As the M2-brane worldvolume extends along more than two directions in \ads{},
a semi-GCA emerges.
It contains a 16 supersymmetric subalgebra
\begin{eqnarray}
\{P_\alpha,~
K_\alpha,~
D,~
J_{05},~
J_{ij},~
J_{a'b'},~
Q_+
\}~.
\end{eqnarray}
The set of generators
$\{H_\alpha,K_\alpha,D,J_{05}\}$
generates the \ads{3} symmetry so(2,2) on the M2-brane worldvolume.
The generators $\{J_{ij},J_{a'b'}\}$ represents
 the rotational symmetry so(4)$^2$ 
in the space transverse to \ads{3} in \adss{7}{4}.

\subsection{16 supersymmetric semi-GCA from osp($8^*|2$)}
We derive a 16 supersymmetric GCA from the $\CN=2$
superconformal algebra osp($8^*|2$).

Firstly, we derive osp($8^*|2$)
from the  $\CN=4$ superconformal algebra
osp($8^*|4$).
 For this, we introduce a 1/2 projector
\begin{eqnarray}
q_+=\frac{1}{2}(1+\Gamma^{789\natural})~,
\end{eqnarray}
which commutes with $p_\pm$,
and require that $\CQ\equiv \CQ q_+$.
Then the $\CN=4$ superconformal algebra reduces to
a $\CN=2$ superconformal algebra.
The fermionic part of the algebra is
\begin{eqnarray}
[\tilde P_\mu,\CQ]&=&-\frac{1}{2}\CQ \Gamma_{\mu 6}p_+~,
~~~
[\tilde K_\mu,\CQ]=+\frac{1}{2}\CQ \Gamma_{\mu 6}p_-~,
~~~
[\tilde D,\CQ]=-\frac{1}{2}\CQ\CI  \Gamma_{6}~,
\nonumber\\
{[}\tilde J_{\mu\nu},\CQ]&=&\frac{1}{2}\CQ \Gamma_{\mu \nu}~,~~~
{[\tilde J_{a'b'},\CQ]}=\frac{1}{2}\CQ  \Gamma_{a'b'}~,
\nonumber\\
\{\CQ^T,\CQ\}&=&
-4C\Gamma^\mu q_+(p_+\tilde K_\mu +p_-\tilde P_\mu )
-2C\Gamma^6 q_+ \tilde D
-C\CI \Gamma^{\mu\nu}q_+\tilde J_{\mu\nu}
+2C\CI\Gamma^{a'b'}q_+ \tilde J_{a'b'}
~.
\label{supralgebra M5 N=2}
\end{eqnarray}
The bosonic part is \bref{conformal} with $\mu=0,\cdots,5$ and
 so(4)
generated by
$\{\tilde J_{a'b'}\}$.
We note that
the commutation relations containing $\tilde J_{a'b'}$
can be rewritten as
\begin{eqnarray}
[\tilde J_I^{(\pm)},\tilde J_J^{(\pm)}]&=&\epsilon_{IJK}\tilde J_K^{(\pm)}~,~~~
[\tilde J_I^{(+)},\CQ]=\frac{1}{2}\CQ \rho_I~,~~~
[\tilde J_I^{(-)},\CQ]=0~,
\nonumber\\
\{\CQ^T,\CQ\}&=&\cdots
+8C\CI q_+ \sum_{I=1,2,3} \rho_I \tilde J_I^{(+)}~,
\end{eqnarray}
with
\begin{eqnarray}
\tilde J_I^{(\pm)}=\left(
\frac{1}{2}(\tilde J_{89}\mp \tilde J_{7\natural}),~
\frac{1}{2}(\tilde J_{79}\pm \tilde J_{8\natural}),~
\frac{1}{2}(\tilde J_{78}\mp \tilde J_{9\natural})
\right)~,~~~
\rho_I=(\Gamma^{89},\Gamma^{79},\Gamma^{78})~.
\end{eqnarray}
It follows that
the set of generators
\begin{eqnarray}
\{\tilde P_\mu,~
\tilde K_\mu,~
\tilde D,~
\tilde J_{\mu\nu},~
\tilde J_I^{(+)},~
\CQ
\}
\end{eqnarray}
forms the superconformal algebra
osp($8^*|2$).
$J_I^{(+)}$ 
generates the R-symmetry su(2)\,$\cong$\,sp(2).

\medskip

As above,
we decompose $\CQ$ as $\tilde Q_\pm=\CQ\ell_\pm$ by introducing 
a pair of projectors $\ell_\pm$ in \bref{ell M5}
and scale generators as in \bref{scaling M5}
and
\begin{eqnarray}
\tilde J_I^{(+)}=J_I^{(+)}~.
\end{eqnarray}
We note that $q_+$ and $\ell_\pm$ commute each other.
Substituting these into the (anti-)commutation relation
of the $\CN=2$ superconformal algebra osp($8^*|2$)
and taking the limit $\omega\to \infty$,
we derive \bref{semi-GCA}, \bref{GCA M5 fermi} and 
\begin{eqnarray}
[J_I^{(+)},J_J^{(+)}]&=&\epsilon_{IJK}J_K^{(+)}~,~~~
[J_I^{(+)},Q_\pm]=\frac{1}{2}Q_\pm \rho_I~,~~~
\nonumber\\
\{Q_+^T, Q_+\}&=&
-4C\Gamma^\alpha\ell_+q_+(p_- P_\alpha
+p_+ K_\alpha)
-2C\Gamma^6 \ell_+q_+ D
-C\CI\Gamma^{\alpha\beta}\ell_+q_+ J_{\alpha\beta}
-C\CI\Gamma^{ij}\ell_+q_+ J_{ij}
\nonumber\\&&
+8C\CI q_+\ell_+ \sum_{I=1,2,3} \rho_I J_I^{(+)}~,
\nonumber\\
\{ Q_+^T,Q_-\}&=&
-4C\Gamma^i\ell_-q_+ ( p_-P_i
+p_+K_i)
-2C\CI\Gamma^{i\alpha}\ell_-q_+ G_{i\alpha}
~.
\label{GCA M5 N=2}
\end{eqnarray}
This is a 16 supersymmetric semi-GCA
\begin{eqnarray}
\{P_\alpha,~
K_\alpha,~
D,~
P_i,~K_i,~G_{i\alpha},~
J_{ij},~
J_{05},~
J_I^{(+)},~
Q_+,~Q_-
\}
~.
\end{eqnarray}
It contains a 8 supersymmetric subalgebra
\begin{eqnarray}
\{P_\alpha,~
K_\alpha,~
D,~
J_{\alpha\beta},~
J_{ij},~
J_I^{(+)},~
Q_+
\}~.
\end{eqnarray}
$\{J_I^{(+)}\}$ acts as the R-symmetry su(2)\,$\cong$\,sp(2).

\section{Summary and Discussions}

Conformal algebra in $(d+1)$-dimensions is 
the \ads{d+2} algebra. 
The IW contraction which derives a GCA
from the conformal algebra in $(d+1)$-dimensions
is equivalent to the IW contraction which derives
a Newton-Hooke string algebra
from the \ads{d+2} algebra.
The string lies along the direction transverse to the boundary.
In other words, a GCA is a boundary realization of a Newton-Hooke string algebra.
From this observation, we have derived
32 supersymmetric GCAs from superconformal algebras,
psu($2,2|4$), osp($8|4$) and osp($8^*|4$).
We have also derived less supersymmetric GCAs
by considering less supersymmetric conformal algebras,
su($2,2|2$), osp($4|4$), osp($2|4$) and osp($8^*|2$).

In this paper we have considered
an \ads{2} string in \adss{5}{5},
an \adss{2}{1} brane in \adss{4}{7}
and an \ads{3} brane in \adss{7}{4}.
There are many other 1/2 BPS brane configurations in \ads{}
spaces as seen in \cite{SY:NH}.
We leave this issue for future studies.
These branes are considered as a probe
which does not affect the background.
The Newton-Cartan-like geometry discussed recently in 
\cite{Bagchi:2009my}\cite{Duval:2009vt}
may have some connections to ours.
It is interesting to clarify this point.

We have chosen the scaling of the supercharge
so that
the resulting super GCA has a 1/2 supersymmetric subalgebra,
which is a supersymmetrization of the base so$(2,1)\times$so($d$) of the bosonic GCA.
As a consistent IW contraction, one may consider another scaling of the supercharges.
In fact, one may scale the supercharges as $\CQ\to \omega^{1/2}\CQ$.
In this case,  the anti-commutator of two $\CQ$s
contains only $P_i$, $K_i$, $G_i$ and $P_{a'}$ for the super GCA 
considered in section 3.1.
So the dynamical supersymmetry disappears.

It is also interesting to look for NR systems which are invariant
under the super GCAs obtained in this paper.
The NR string action \cite{GGK}\footnote{
See \cite{SY:NH} for the NR brane actions.},
which has the symmetry of the super Newton-Hooke algebra of a string,
might give some hint for this.

In this paper, we have not discussed
the central extensions of the GCA
and the infinite-dimensional
extensions of the GCA
\cite{Bagchi:2009my}.
For the centrally extended GCA, the exotic GCA \cite{Lukierski:2005xy},
we can derive it as an expansion \cite{expansion,deAzcarraga} of the conformal algebra.
In addition, one may find non-central extensions of a (super) GCA
by this method.
We hope to report on this issue in future.

We hope that our results may be useful in the future studies in AdS/CFT.

\subsection*{Note added}
The supersymmetric extensions of GCA
derived in section 3
were found independently
by J. A. de Azcarraga and J. Lukierski
in \cite{dAL}.

\section*{Acknowledgment}

The authors would like to thank 
Kentaroh Yoshida
for
useful discussion. This work
was supported by the
Grant-in-Aid for Scientific Research (19540324) from the Ministry of
Education, Science and Culture, Japan.


\end{document}